\documentclass[twocolumn]{aastex631}
\let\tablenum\relax
\usepackage{amsmath}
\usepackage{tikz}
\usetikzlibrary{matrix}
\usetikzlibrary{graphs}
\usetikzlibrary{chains,scopes}
\usetikzlibrary{arrows.meta, automata, positioning, fit}
\newcommand{\dark}{\texttt{dark}~}
\newcommand{\light}{\texttt{light}~}

\submitjournal{PSJ}

\begin{document}

\title{Lunar geochemistry from X-ray line flux ratios using CLASS on Chandrayaan 2}

\correspondingauthor{R. Kumar}
\email{kumar0428@umn.edu}

\author[0009-0008-6428-7668]{R. Kumar}
\affiliation{Department of Aerospace Engineering, Indian Institute of Technology Bombay, Powai, Mumbai 400076, India}
\affiliation{School of Physics and Astronomy, University of Minnesota, Minneapolis, MN 55455, USA}

\author[0009-0005-5080-0107]{Y. Rai}
\affiliation{Department of Physics, Indian Institute of Technology Bombay, Powai, Mumbai 400076, India}

\author[0009-0007-1358-1878]{S. Srijan}
\affiliation{Department of Computer Science and Engineering, Indian Institute of Technology Bombay, Powai, Mumbai 400076, India}

\author[0009-0002-7087-4750]{A. Bansal}
\affiliation{Department of Computer Science and Engineering, Indian Institute of Technology Bombay, Powai, Mumbai 400076, India}

\author[0009-0001-1434-4600]{Ameya V Singh}
\affiliation{Department of Computer Science and Engineering, Indian Institute of Technology Bombay, Powai, Mumbai 400076, India}

\author[0009-0000-6358-0581]{A. Kumar}
\affiliation{Department of Physics, Indian Institute of Technology Bombay, Powai, Mumbai 400076, India}

\author[0009-0000-1053-9470]{H. Mhatre}
\affiliation{Department of Civil Engineering, Indian Institute of Technology Bombay, Powai, Mumbai 400076, India}

\author[0009-0004-9984-4138]{M. Goyal}
\affiliation{Department of Physics, Indian Institute of Technology Bombay, Powai, Mumbai 400076, India}

\author[0000-0000-0000-0000]{S. Swain}
\affiliation{Department of Physics, Indian Institute of Technology Bombay, Powai, Mumbai 400076, India}

\author[0009-0007-3245-3820]{S. Patidar}
\affiliation{Department of Computer Science and Engineering, Indian Institute of Technology Bombay, Powai, Mumbai 400076, India}

\author[0009-0005-2987-0688]{Aditya P Saikia}
\affiliation{Department of Physics, Indian Institute of Technology Bombay, Powai, Mumbai 400076, India}

\author[0009-0002-9447-7346]{A. Ahmad}
\affiliation{Department of Computer Science and Engineering, Indian Institute of Technology Bombay, Powai, Mumbai 400076, India}

\author[0000-0001-9199-4925]{S. Narendranath}
\affiliation{U R Rao Satellite Centre, ISRO, Bengaluru, India}
\affiliation{Raman Research Institute, Bengaluru, India}
\author[0000-0001-8525-4867]{Netra S Pillai}
\affiliation{U R Rao Satellite Centre, ISRO, Bengaluru, India}

\author[0000-0002-5700-282X]{R. Kashyap}
\affiliation{Department of Physics, Indian Institute of Technology Bombay, Powai, Mumbai 400076, India}

\author[0000-0002-6112-7609]{V. Bhalerao}
\affiliation{Department of Physics, Indian Institute of Technology Bombay, Powai, Mumbai 400076, India}

\begin{abstract}
Global lunar chemical maps are essential for understanding the origin and evolution of the Moon, its surface characteristics, and its potential for resource extraction. Lunar elemental abundance maps have been derived using X-ray and gamma ray spectroscopy previously but are limited in coverage or have coarse spatial resolution. Here we used X-ray fluorescence line intensity of O, Mg, Al, Si, Ca and Fe derived from five years of data from the Chandrayaan-2 Large Area Soft X-ray Spectrometer (CLASS) to generate global O/Si, Mg/Si, Al/Si, Mg/Al, Ca/Si and Fe/Si line intensity ratio maps at a resolution of 5.3~km/pixel. We have developed an independent data analysis methodology for CLASS, based on open source Python packages.  Our analysis shows that the Mg/Al map best represents the geochemical differences between the major terranes, consistent with the findings of the Apollo 15 and 16 X-ray Fluorescence Spectrometer (XRS) maps. We have also shown a good correlation of the line intensity ratios with the abundance ratios from CLASS using published elemental abundance maps.  Further, we apply Gaussian mixture models to the Mg/Si vs Al/Si density maps to map geochemically distinct regions on the Moon that could be of interest for future investigations.
\end{abstract}


\section{Introduction} \label{sec:intro}
The lunar surface has been classified into three main crustal terranes \citep{2000JGR...105.4197J} based on global surface geochemistry. These three large units, namely the Procellarum KREEP Terrane (PKT), Feldspathic Highlands Terrane (FHT), and the South Pole--Aitken Terrane (SPAT) have uniquely different evolutionary histories.  Although quantitative elemental abundance maps, such as those obtained from X-ray and gamma-ray spectroscopy, are necessary to model the evolution of the distinct compositions, their usage has been mainly limited to these broad terranes because of the coarse spatial resolutions of such lunar elemental maps. Future lunar missions that explore a small region of interest and sample returns would require more detailed mapping to identify potential target sites. 

X-ray fluorescence (XRF) spectroscopy is a well established technique in planetary remote sensing of airless bodies \citep{1972Sci...175..436A, adler1972, 2009Icar..200..129L, 2009P&SS...57..744S, 2011Icar..214...53N, 2011Sci...333.1847N,2014Icar..229..254W,2016RAA....16....4D} for elemental mapping.  XRF is triggered when high-energy photons ionize inner-shell electrons, causing the emission of X-ray fluorescent photons unique to each element. The intensity of the element's spectral lines is proportional to its abundance. However, determining elemental abundances using the XRF spectra  depends on several factors, such as the solar spectrum, the geometry of the observation, the grain size , the matrix of elements present, and their possible vertical layering \citep{1966JaJAP...5..886S, 1979PhDT.......224C,1997JGR...10216361C,1997LPI....28.1039O,2008EP&S...60..293M,2008EPS...60..283O,2008Icar..198..408N,2009AdSpR..44..313N,2011P&SS...59.1393W, ATHIRAY2013188, 2013PSS...89..183A}. The uncertainties in each of these factors are propagated to the elemental abundance maps, which limits their accuracy. 

Chandrayaan-2 Large Area Soft X-ray Spectrometer \citep[CLASS;][]{vatedka2020chandrayaan} measures lunar X-ray spectra with a nominal spatial resolution of $12.5$~km~$\times$~$12.5$~km from a circular polar orbit of $100 \pm 20$~km. The lunar X-ray line flux is strongly dependent on solar activity, and often several spectra spanning larger ground pixel tracks have to be co-added to increase statistics. \citet{2024Icar..41015898N} derived lunar elemental maps at a spatial resolution of $150$~km~$\times$~$12.5$~km from CLASS. Although spectral measurements were nearly global, abundance maps were limited in coverage as only the best spectral fits with good statistics could be used. 

X-ray line flux ratios are representative of geochemistry, as shown by XRF experiments, such as lunar measurements on Apollo 15 \citep{1972Sci...175..436A} and 16 \citep{adler1972}, as well as measurements for Asteroid 433 Eros \citep{2009Icar..200..129L} and Mercury \citep{weider2015, 2020Icar..34513716N}.

Here, we used five years of data from CLASS to generate global elemental line intensity maps of O/Si, Mg/Si, Al/Si, Mg/Al, Ca/Si and Fe/Si. Using partially overlapping tracks from multiple passes, we bin data into a uniform grid of $5.3$~km~$\times$~$5.3$~km, which is higher than the inherent spatial resolution of CLASS.

This paper is organized as follows. In \S\ref{sec:methodolgy}, we discuss the new independent pipeline that we developed for analysis of the CLASS L1 data.
The ratio maps are presented in Section \S\ref{sec:mapping} followed by a discussion in \S\ref{sec:lunar-chemistry}. 
The source data was downloaded from the ISRO Science Data Archive\footnote{\url{https://pradan.issdc.gov.in/ch2/}}, and our codes and data products are publicly available on GitHub\footnote{\url{https://github.com/ravioli1369/selenomapper.git}} and Zenodo\footnote{\url{https://doi.org/10.5281/zenodo.16726150}}.

\section{Methodology}\label{sec:methodolgy}

CLASS L1 data consist of energy-calibrated lunar X-ray spectra at a cadence of $8$~s, which corresponds to a ground pixel of $12.5$~km~$\times 12.5$~km on the lunar surface from an altitude of $100$~km. We used L1 data from November 2019 to May 2025 for our analysis. Although the total number of spectra is close to 7~million, not all of them have a lunar XRF signal due to the dependence on solar activity. We therefore generated algorithms for filtering out lunar X-ray spectra where XRF signals are detected. Although the XRF signals are line emissions, there is a continuum arising from the charged particles in the lunar environment as well as scattered X-rays. The appropriate background has to be subtracted from the spectra to confirm the detection of XRF lines and ensure that we obtain accurate line ratios.

\subsection{Background Estimation}\label{sec:bkgest}
The background in CLASS detectors arises from Galactic cosmic rays, solar energetic particles, and geotail particles. We estimate this charged particle background from the night-side observations. 
CLASS mainly collects data on the sunlit portion of the Moon, which we refer to as \light data. It also collects about five minutes of data on the night-side of the moon, which we refer to as \dark data, and it can help in estimating the instrument background.
We note that the orbiter is not fully in the Moon's shadow throughout the \dark phase, because of its altitude $h$ above the lunar surface, which can lead to improper modeling of the background. For example, we note the presence of a strong Al-K$\alpha$ and a weak Cu-K$\alpha$ line in the \dark spectra at the time of transition from \light to \dark. These lines are excited from the optical light-blocking filters as solar X-rays illuminate the filters and the copper collimator at grazing incidence, before the orbiter is fully occulted by the Moon.
We filter the \dark data using the following condition:
\begin{equation}
R_M (\csc \theta - 1) \geq h ,
\end{equation}
where $R_M$ is the radius of the Moon, $\theta$ is the solar angle (the angle between the surface normal and sun vector), and $h$ is the orbiter altitude given in the FITS data headers. In addition, \dark spectra with strong instrumental lines were also filtered out.

We use the spectra measured in this filtered subset of \dark spectra and subtract it from the \light spectra before estimating elemental abundances. 
Since the background in the \light and \dark stages will not be exactly identical due to the missing direct solar contribution, we apply a first-order correction for this as discussed in \S\ref{sec:xrfdet}. The Al-K$\alpha$ and a weak Cu-K$\alpha$ seen at the start and end of the \dark data are not expected to be present in the \light data as sun rays cannot directly reach the optical filters due to the spacecraft orientation.

\begin{figure}[h!]
    \centering
    \includegraphics[width=0.9\linewidth]{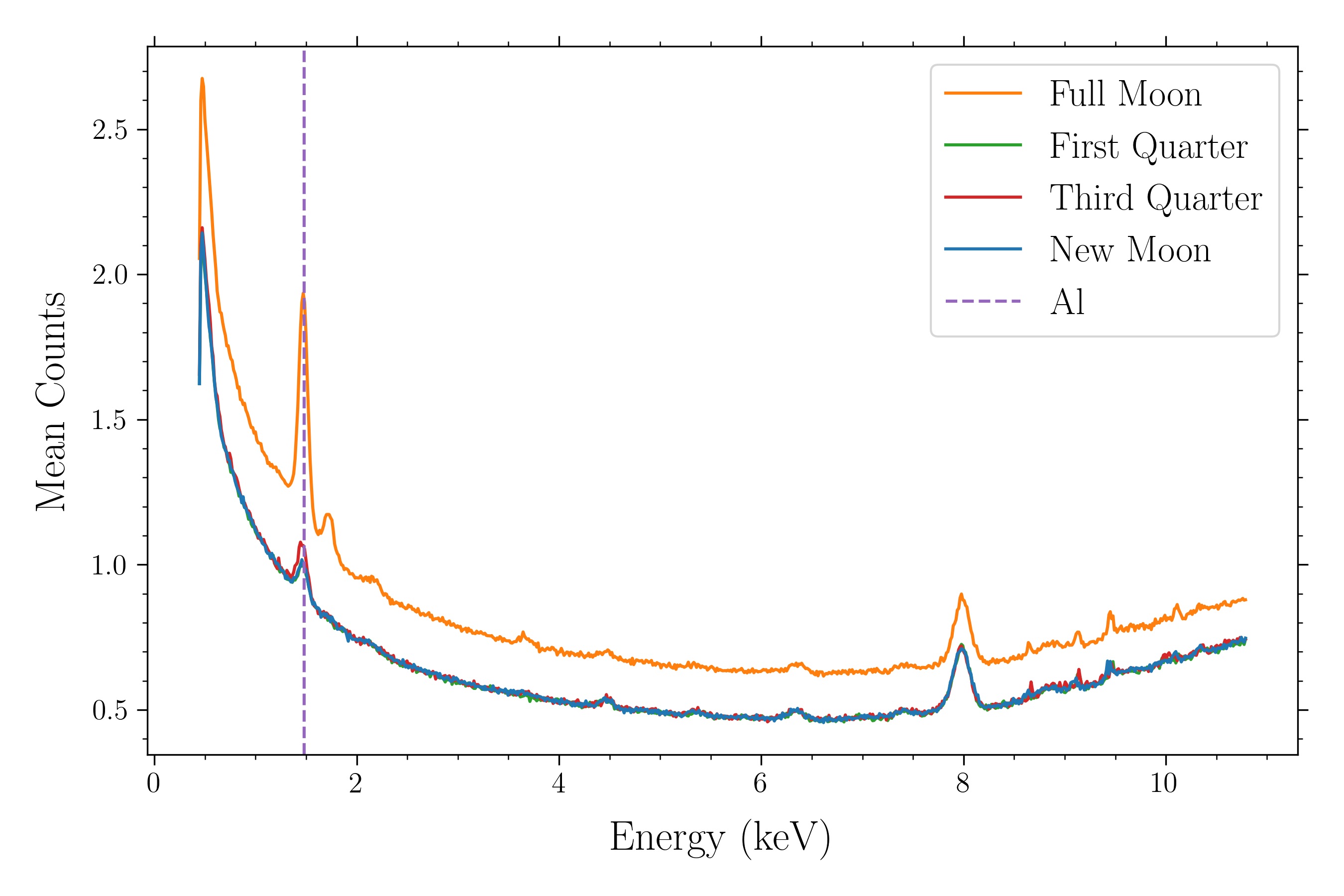}
    \caption{Mean background spectrum from one year (2020) of CLASS data measured in the night side of the orbit at different lunar phases is shown.  During geotail passes, the mean background is higher with strong instrumental lines of Al-K$\alpha$, Cu-K$\alpha$, and a weak Si-K$\alpha$.}
    \label{fig:bkg}
\end{figure}

We also note that the amount of \dark data per lunar day that is occulted is a very small fraction ($\approx 14\%$) of the total \dark data, and there are often no occulted data for some orbits. We take a dynamic yearlong phase-wise background (Figure \ref{fig:bkg}), where we define the phase as $\pm 3.5$~days around the Full Moon (FM), New Moon (NM), First Quarter (FQ) and Third Quarter (TQ) to account for the phase-dependent background variations. The pipeline calculates and stores background spectra for each occurrence of these phases. When analyzing \light data for abundances, the pipeline uses the background measurements from a period of $\pm 6$~months from the Earth day of the \light observation. Sample background spectra for the four phases for 2020 are shown in Figure~\ref{fig:bkg}.

\subsection{XRF detection and analysis}\label{sec:xrfdet}
Based on the energy range of CLASS, we analyze spectra to search for XRF lines of Oxygen-K$\alpha$ ($0.52$~keV), Sodium-K$\alpha$ ($1.04$~keV), Magnesium-K$\alpha$ ($1.25$~keV), Aluminum-K$\alpha$ ($1.48$~keV), Silicon-K$\alpha$ ($1.74$~keV), Calcium-K$\alpha$ ($3.69$~keV) and $\beta$ ($4.01$~keV), Titanium-K$\alpha$ ($4.51$~keV), Chromium-K$\alpha$ ($5.41$~keV), Manganese-K$\alpha$ ($5.89$~keV) and Iron-K$\alpha$ ($6.4$~keV) and L ($0.7$~keV).

We estimate the detection threshold based on the variation in background counts (from the set of selected \dark spectra) over time at the energy of each line. Given the energy resolution of CLASS, widths of typical lines, and the spacing between the lines listed above, we opted to measure background counts from a 0.25~keV window centered at each line energy. Consider a transition line $T$ for some element $X$ with an energy $E_T^X$. The total background counts in the $E_T^X \pm 0.125$~keV window at a lunar phase $P$ are denoted by $\text{Counts}^X_T(\mathcal{B}_P)$, and the peak amplitude within that window is denoted by $\text{Amp}^X_T\left(\mathcal{B}_P\right)$. Using the year-long set of dark measurements, we now estimate the standard deviations of the counts, $\sigma\left(\text{Counts}^X_T \left(\mathcal{B}_P\right)\right)$; and of the amplitude, $\sigma\left(\text{Amp}^X_T \left(\mathcal{B}_P\right)\right)$, which are used in detection criteria.

We observe the standard deviation of the background Counts during different moon phases follows the trend $\sigma\left(\mathcal{B}_{FM}\right) \gg \sigma\left(\mathcal{B}_{FQ}\right) \approx \sigma\left(\mathcal{B}_{TQ}\right) > \sigma\left(\mathcal{B}_{NM}\right) $, further reinforcing the choice of using a phase-dependent background.

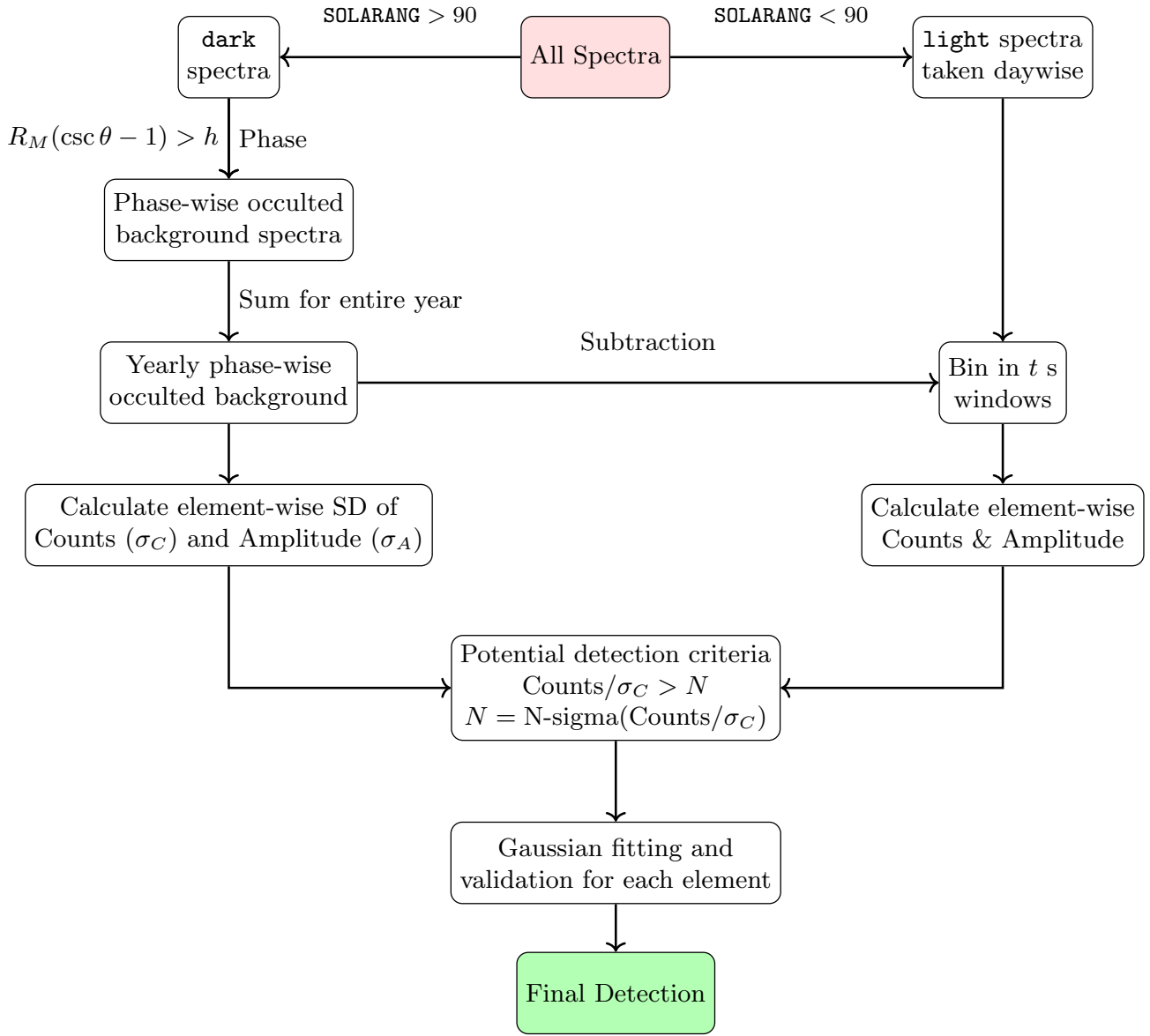
\begin{figure*}[hp]
    \centering
\resizebox{0.98\textwidth}{!}
{ 
\begin{tikzpicture}[
    normal/.style={draw, rectangle, rounded corners, align=center, minimum height=1cm},
    every edge/.style={draw, -stealth},
    label/.style={draw=none, fill=none, align=center, minimum height=1cm}
  ]
    \def\edgewidth{0.8}
  \node[normal] (A) at (-1, -2) {Phase-wise occulted \\ background spectra};
  \node[normal] (B) at (-1, 0) {\dark \\ spectra};
  \node[normal, fill=pink!50] (C) at (3.5, 0) {All Spectra};
  \node[normal] (D) at (8.5,0) {\light spectra \\ taken daywise};
  \node[normal] (F) at (-1, -4) {Yearly phase-wise \\ occulted background};
  \node[normal] (G) at (8.5,-4) {Bin in $t$~s \\ windows};
  \node[normal] (H) at (-1,-5.75) {Calculate element-wise SD of \\ Counts ($\sigma_C$) and Amplitude ($\sigma_A$)};
  \node[normal] (I) at (8.5,-5.75) {Calculate element-wise \\ Counts \& Amplitude};
  
  \node[normal] (J) at (3.75,-7.75) {Potential detection criteria \\ $\text{Counts}/\sigma_C > N$ \\ $N = \text{N-sigma}(\text{Counts}/\sigma_C)$};
  \node[normal] (K) at (3.75,-9.9) {Gaussian fitting and \\validation for each element};
  \node[normal, fill=green!30] (L) at (3.75,-11.5) {Final Detection};
  
  \draw[->, line width = \edgewidth] (B) -- node[label, midway, left] {$R_M(\csc \theta - 1) > h$} (A);
  \draw[->, line width = \edgewidth] (C) -- node[label, midway, above] {\footnotesize{$\texttt{SOLARANG} > 90$}} (B);
  \draw[->, line width = \edgewidth] (C) -- node[label, midway, above] {\footnotesize{$\texttt{SOLARANG} < 90$}} (D);
  \draw[->, line width = \edgewidth] (B) -- node[label, midway, right] {Phase} (A);
  \draw[->, line width = \edgewidth] (A) -- node[label, midway,right] {Sum for entire year} (F);
  \draw[->, line width = \edgewidth] (F) -- node[label, midway, above] {Subtraction} (G);
  \draw[->, line width = \edgewidth] (D) -- node[label, midway, right] {} (G);
  \draw[->, line width = \edgewidth] (F) -- node[label, midway, right] {} (H);
  \draw[->, line width = \edgewidth] (G) -- node[label, midway, right] {} (I);
  \draw[->, line width = \edgewidth] (H) |- (J);
  \draw[->, line width = \edgewidth] (I) |- (J);
  \draw[->, line width = \edgewidth] (J) -- node[label, midway, right] {} (K);
  \draw[->, line width = \edgewidth] (K) -- node[label, midway, right] {} (L);
  \end{tikzpicture}
    } 
\caption{Flow chart of detection algorithm}
\label{fig:detection-algo}
\end{figure*}

The instrument acquires data in 8~s exposures. Following ~\citet{NARENDRANATH2024115898}, we group the data into 96~s bins for our analysis. Defining $\mathcal{L}_t$ as the \light data corresponding to the bin starting at time $t$, we obtain the integrated counts in each line $T$ as $\text{Counts}^L_T\left(\mathcal{L}_t\right)$ using a similar 0.25~keV window centered on the line. We have to correct this for two effects. The first is simply the appropriate background for the corresponding line in that lunar phase $\mathcal{B}_P$. The second is the difference between the \light and \dark side spectra due to the scattered solar X-rays or particle interactions in the \light data, discussed in \S\ref{sec:bkgest}. By investigating the spectra, we find that this contribution is not strongly dependent on energy --- hence we simply model it as the difference between the median counts in background and \light spectra in the relatively line-free region of 2 -- 7~keV. Thus, the corrected line data are given by $\mathcal{L}^s_t = \left(\mathcal{L}_t - \mathcal{B}_P\right) - \mathrm{median}\left(\mathcal{L}_t - \mathcal{B}_P\right)$. 

We now look for statistically significant detections of fluorescence lines in the normalized spectra, defined as
\begin{equation}
    \mathcal{N}_t = \frac{\text{Counts}^X_T(\mathcal L^s_t)}{\sigma\left(\text{Counts}_T^X\left(\mathcal{B}_P\right)\right)} \label{eq:norm_l}
\end{equation}
where the subscript $t$ symbolizes the time stamp of observation during that orbit.
Nominally, a 5-$\sigma$ detection here would be defined as $\mathcal{N}_t > 5$. However, since the background measurement is for a long duration, it can have significant variations, increasing the denominator in Equation~\ref{eq:norm_l}. As a consequence, this threshold may not be crossed in periods of low solar activity. In such periods, we apply a different threshold based on the median and standard deviation of the light curve of the line flux during that orbit, given by:
\begin{equation}
   \mathcal{N}_t \geq \text{median}(\mathcal{N}_t) + 5 \sigma(\mathcal{N}_t),
    \label{eq:detection-criteria1}
\end{equation}

Thus, the final detection criterion applied for each element is as follows:
\begin{equation}
   \mathcal{N}_t \geq min \left(5,~\text{median}(\mathcal{N}_t) + 5 \sigma(\mathcal{N}_t) \right)
    \label{eq:detection-criteria}
\end{equation}

Data satisfying Equation~\ref{eq:detection-criteria} are now analysed to measure line strengths.
We fit the spectral lines with a Gaussian function $(\mathcal{G})$ to obtain the integrated counts under each line. 
We use specutils's \texttt{fit\_lines} function for fitting Astropy's \texttt{Gaussian1D} object to the spectrum object created using specutil's \texttt{Spectrum1D} class. At this stage, we found that we obtained some spurious measurements due to outliers in the spectra. Based on the expected values of the parameters of the Gaussian fit, apply another set of criteria (Equations~\ref{eq:detection-criteria2a}--\ref{eq:detection-criteria2c}) to eliminate false detections:

\begin{align}
    \frac{\text{Amp}_T^X\left(\mathcal{G}\left(\mathcal{L}_t^s\right)\right)}{\sigma \left(\text{Amp}_T^X\left(\mathcal{B}_P\right)\right)} & \ge min (5,~med + 5 \sigma) , \label{eq:detection-criteria2a} \\
    \mu\left(\mathcal{G}\left(\mathcal{L}_t^s\right)\right) - E^{X}_T & \in [-0.05, 0.05], \label{eq:detection-criteria2b}\\ 
    \sigma\left(\mathcal{G}\left(\mathcal{L}_t^s\right)\right) & \in [0.05, 0.1]. \label{eq:detection-criteria2c}
\end{align}

Here, $\text{Amp}_T^X\left(\cdot\right)$ denotes the amplitude, making Equation~\ref{eq:detection-criteria2a} the amplitude analog of the counts-based Equation~\ref{eq:detection-criteria}. Equation~\ref{eq:detection-criteria2b} requires the Gaussian fit mean $\mu$ to be close to the known line energy $E_T^X$, and Equation~\ref{eq:detection-criteria2c} requires the line width to be consistent with expectations.

\begin{figure}[ht]
    \centering
    \includegraphics[width=\linewidth]{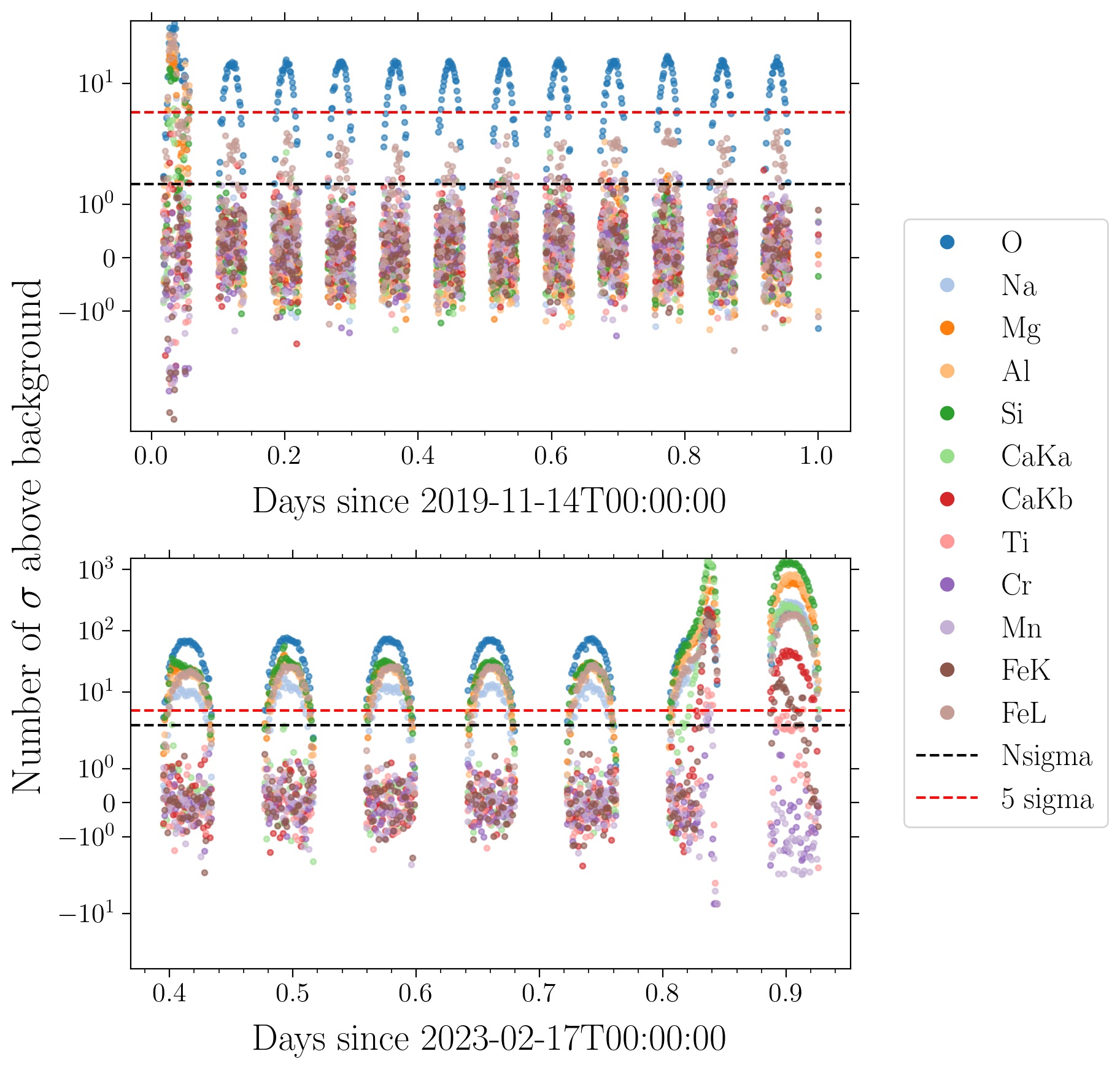}
    \caption{Light curves during a  typical day during solar minima (top) and solar maxima (bottom).}
    \label{fig:lc_samples}
\end{figure}

Figure \ref{fig:detection-algo} illustrates the steps described here.  There are 238,817 XRF line detections that pass our set of criteria (Equations~\ref{eq:detection-criteria2a}--\ref{eq:detection-criteria2c}) from November 2019 to May 2025. As solar activity increases, there are more detections, especially for elements with higher atomic numbers. Typical light curves with detection thresholds are shown in Figure \ref{fig:lc_samples}. After processing the data, we observed that the signal-to-noise ratio was low for some of the line ratios. Hence, calculations for Mg/Si, Fe$_L$/Si, O/Si, and Ca-K$\alpha$ lines were repeated using  296~s binning.

\subsection{Propagation of uncertainties}

In the observed counts, we have contributions from the background and the source. By using propagation of errors we can obtain the variance in the background subtracted observed counts. We also assume poisson statistics for both the background and source photons for our uncertainty calculcations.

Suppose that there are $N_\mathcal{L}$ spectra to be added to create the binned light spectra at time $t$ $\left(\mathcal{L}_t\right)$, and $N_\mathcal{B}$ spectra that are added to make up the background spectrum $\left(\mathcal{B}\right)$. Then, by adding up the variances, scaled by the number of observations $\left(N_\mathcal{L},N_\mathcal{B}\right)$, of $\left(\mathcal{L}_t\right)$ and $\left(\mathcal{B}\right)$, we get:

\begin{equation}
       std(\mathcal{L}^s_t) = \sqrt{\frac{\mathcal{L}_t}{N_\mathcal{L}} + \frac{\mathcal{B}}{N_\mathcal{B}}}
    \label{eq:errors2}
\end{equation}

We pass $std(\mathcal{L}^s_t)$ obtained from equation \ref{eq:errors2} to the $\texttt{Spectrum1D}$ object, which then weighs each data point accordingly. These weights are used to decide the optimal fit and are propagated in the fitted parameters.

\section{Elemental ratio maps}\label{sec:mapping}

The maps are produced as a GeoTIFF file using the standard coordinate reference system (CRS) EPSG:4326. For calculating the ratios, we filter out the detections based on the criteria described in Equations~\ref{eq:detection-criteria2a}--\ref{eq:detection-criteria2c}, and calculate the area under the Gaussian fits for various elements. We define the ratio $\mathcal{R}^M_L$ of element $M$ with element $L$ as:
\begin{equation}
    \mathcal{R}^{M}_{L} = \dfrac{A_M \sigma_M}{A_L \sigma_L} ,
\end{equation}
where $A_M$, $A_L$ and $\sigma_M$, $\sigma_L$ are the amplitudes and standard deviations of the Gaussian fits to the $M$ and $L$ spectral lines respectively.
We take all ratios with the base element as Silicon and derive the uncertainties from the covariance matrix given by \texttt{fit\_lines} of the gaussian fits. We take the uncertainty weighted average and propagate uncertainties for each pixel. Each pixel in the map corresponds to an area of $\sim$$5.3\times 5.3$~km on the lunar surface. 

\begin{figure*}
    \gridline{\fig{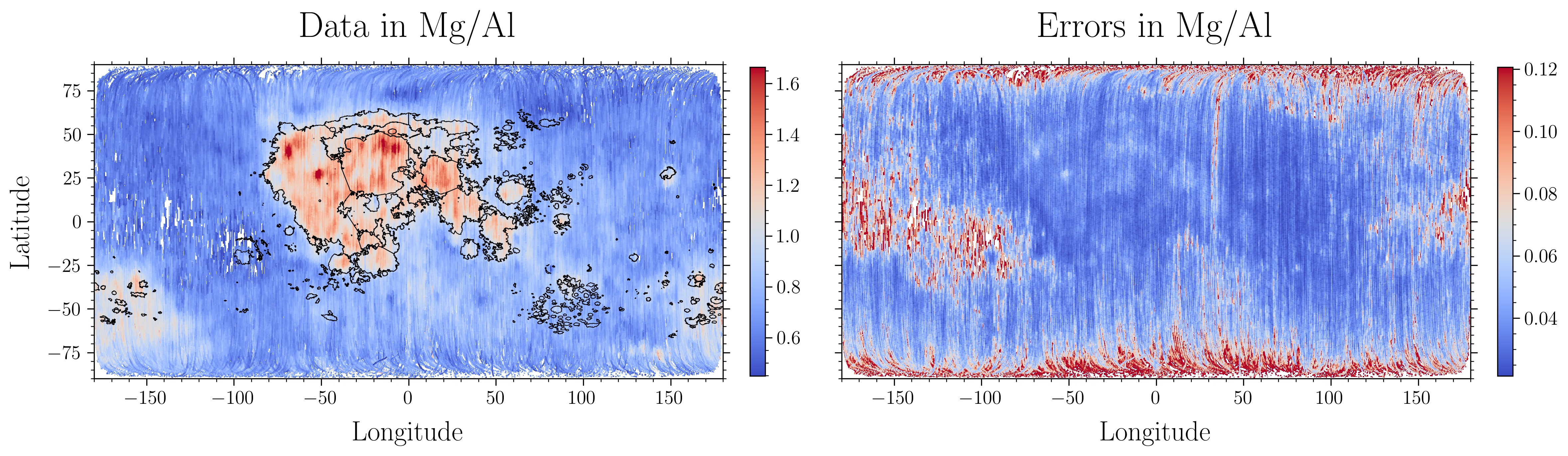}{0.96\textwidth}{}}
    \vspace{-25pt}
    \gridline{\fig{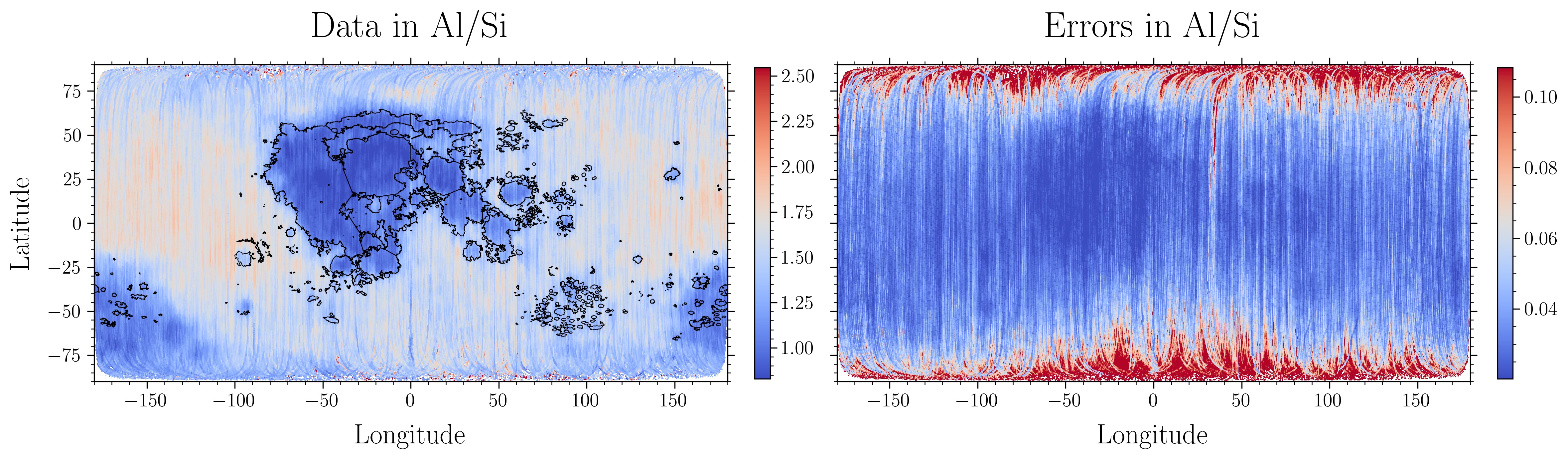}{0.96\textwidth}{}}
    \vspace{-25pt}
    \gridline{\fig{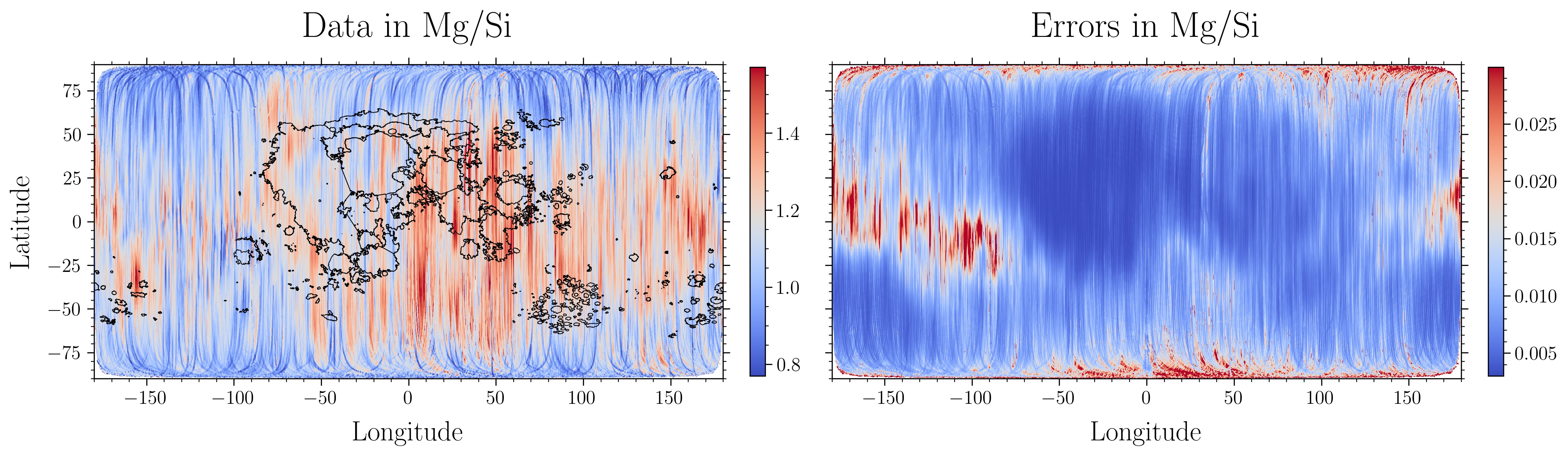}{0.96\textwidth}{}}
    \vspace{-25pt}
    \gridline{\fig{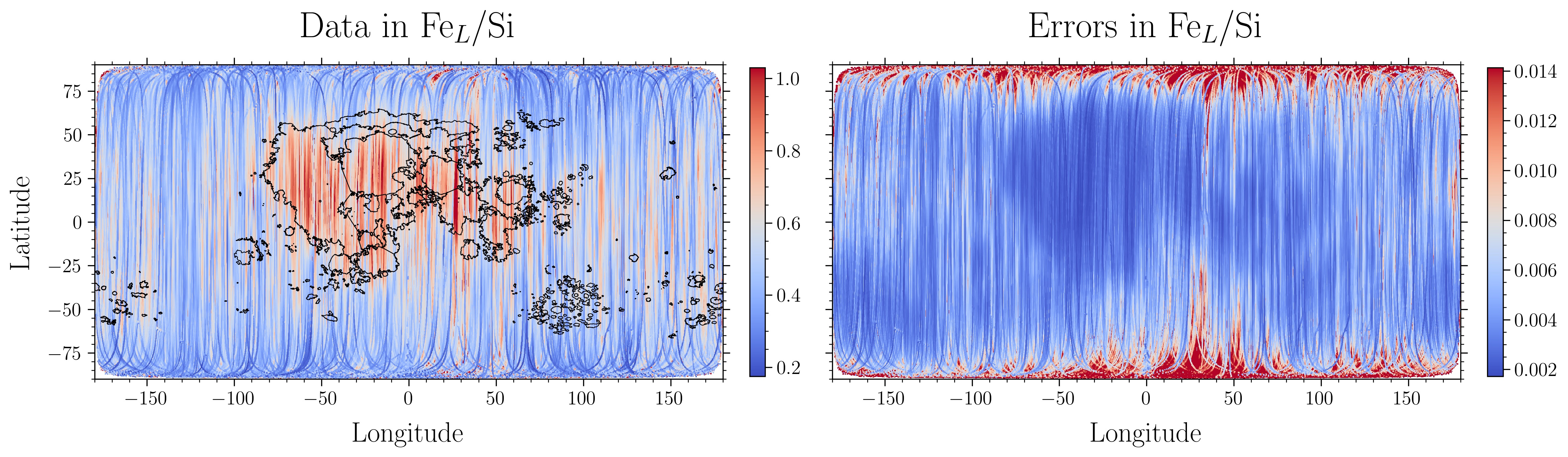}{0.96\textwidth}{}}
    \vspace{-25pt}
    \caption{Lunar XRF line intensity ratio maps from CLASS data. Mg/Al and Al/Si best represents the distinction between the major lunar terranes. High Mg/Si values are not restricted to Mare while Fe$_L$/Si is higher in mare but not so evidently in SPA. Black lines are Mare boundaries from LROC data. As mentioned in \S\ref{sec:xrfdet}, Mg/Al and Al/Si maps were created using 96~s data bins, while the Mg/Si and Fe$_L$/Si maps used 296~s binning.}
    \label{fig:maps1}
\end{figure*}

Figure \ref{fig:maps1} shows the Mg/Al, Al/Si, Mg/Si,  Fe$_L$/Si and Ca/Si maps which have global coverage. Mg/Al and Al/Si data best represent the known distinction in chemistry between Mare, highlands, and the South Pole Aitken Terrane (SPAT). The uncertainties are higher in the polar regions as the signal is weaker there because of the grazing solar incidence angles. For Mg/Al and Mg/Si, uncertainties are also higher in the highland terrane.

As a visual guide, we have added Mare boundaries on all maps. These boundaries are based on the digitization of lunar mare boundaries from Lunar Reconnaissance Orbiter Camera (LROC) Wide Angle Camera (WAC) basemap Shapefile, and downloaded from the Lunar Reconnaissance Orbiter Camera Science Operations Center website\footnote{\url{https://data.lroc.im-ldi.com/lroc/view_rdr/SHAPEFILE_LROC_GLOBAL_MARE}.}.

\begin{figure}
    \includegraphics[width=\linewidth]{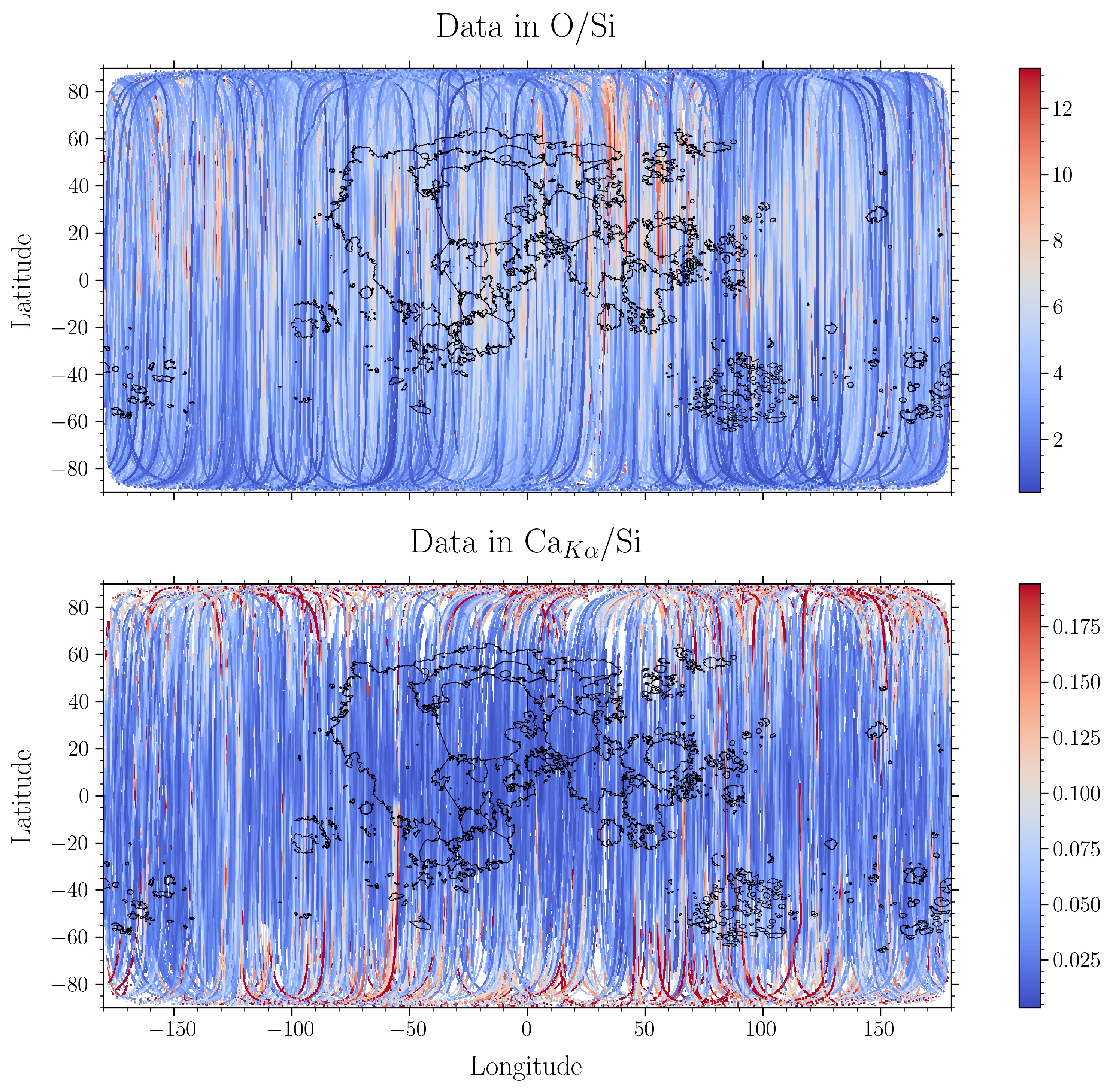}
    \caption{Lunar XRF line intensity ratio maps of O/Si and Ca-K$\alpha$/Si from CLASS data binned at 296~s. Black lines are Mare boundiares from LROC data. Oxygen is not expected to vary much across the lunar surface, which is indeed observed in the O/Si map. On the other hand, the expected difference in terranes are not observed in Ca-K$\alpha$ map. }
    \label{fig:maps2}
\end{figure}

O/Si and Ca-K$\alpha$/Si maps shown in Figure \ref{fig:maps2} do not distinguish between the terranes. Oxygen is expected to be in the narrow abundance range of 40-45 wt \% across the Moon, but it is difficult to validate the O/Si map, as there have been no such global measurements previously. The difference in XRF line energies between Si and Ca would lead to a better excitation of Ca lines for harder solar spectra, which makes the Ca/Si line ratio not suitable for representing composition. Na/Si, Ti/Si, Cr/Si and Fe-K$\alpha$/Si have lower coverage and are not discussed here. 

\subsection{Correlation with abundances}\label{sec:corr_abund}
X-ray line flux ratios (rather than abundance in wt\%) have been effectively utilized for surface compositional studies of planetary bodies in the past \citep{2012JGRE..117.0L05W, Nittler2018}. 
This is especially important for planetary objects, where ground truths are not available and the results are entirely based on modeling the XRF spectra. Here we used the Mg, Al, Si, and Fe wt\% published maps from CLASS \citep{NARENDRANATH2024115898} to demonstrate the validity of elemental ratios for geochemical studies. 

\begin{figure}
    \centering
    \includegraphics[width=\linewidth]{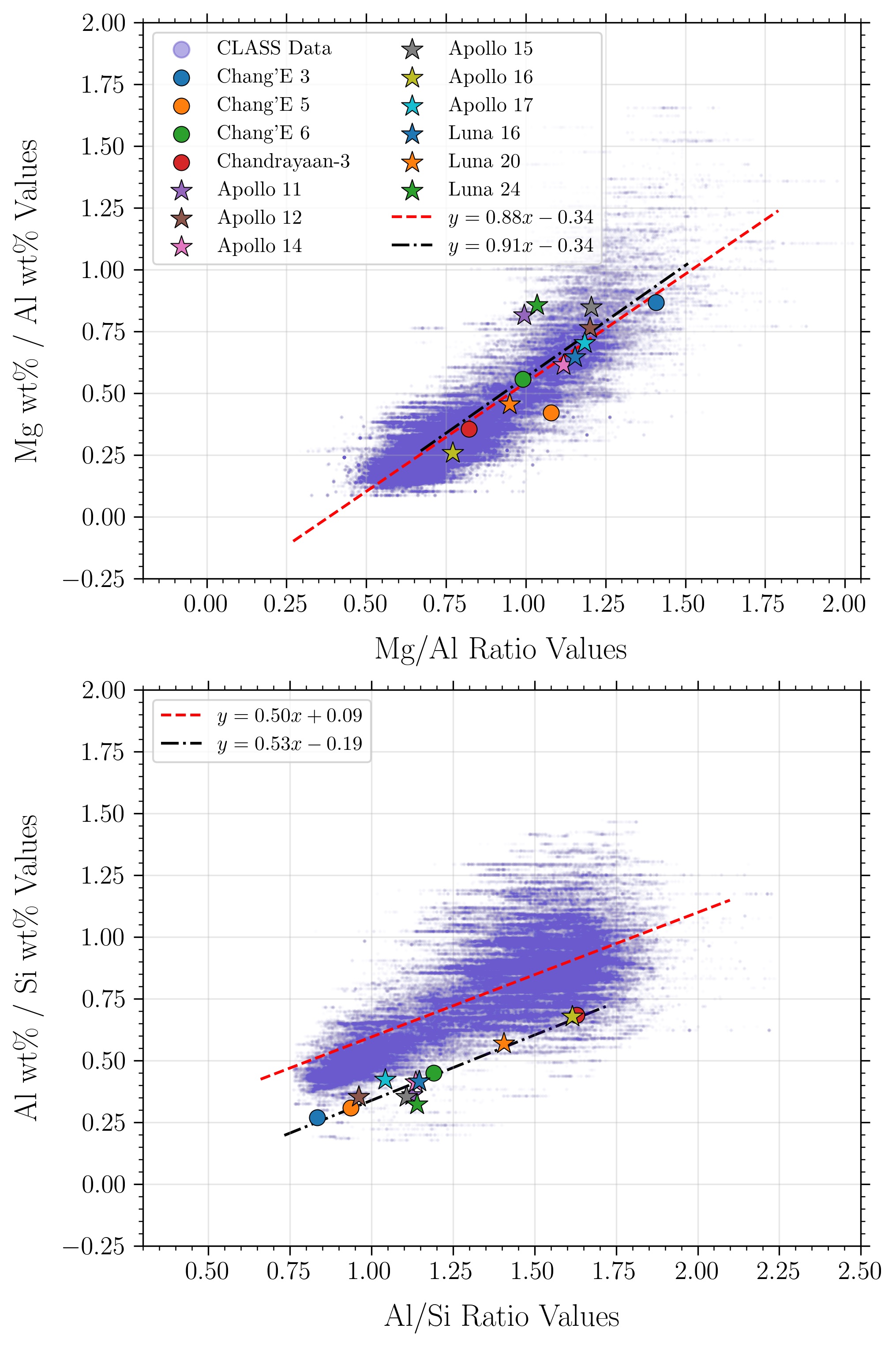}
    \caption{Comparing XRF line intensity ratios with elemental abundance data. Purple circles points denote inferred elemental weight ratios from \citet{NARENDRANATH2024115898} as compared with the fluorescence strength ratio. The star symbols denote actual measurements from various missions. We see a good linear correlations in the measurements. The red dashed lines denote the linear fit to CLASS data, while the black dash-dotted line denotes a linear fit to elemental weight ratio data from various missions.}
    \label{fig:ratio_abund}
\end{figure}

Figures \ref{fig:ratio_abund} shows a very good correlation of Mg/Al and Al/Si ratios with the corresponding wt\%.

We used the line ratios corresponding to the sample return sites for our comparison with the abundances of the returned soils from Apollo, Luna \citep{meyer2005lunar}, Chang'E 3 \citep{Ng2023}, Chang'E 5 \citep{Zong2022} and Chandrayaan-3 APXS \citep{Vadawale2024} and Chang'E 6 \citep{10.1093/nsr/nwae328}. 

Mg/Al line ratios for both CLASS global measurements and return sample sites are very similar. But Al/Si wt\% is higher for CLASS global measurements than for return sample sites for a given line ratio. 
  
The difference probably arises because Si wt\% from CLASS has a larger range of values (and higher uncertainty) with a mode of $\approx 18 $ wt\%, than in the returned samples where the range is 16-23 $wt\%$.

Mg/Al line ratio values $\mathcal{R}^\text{Mg}_\text{Al} \lessapprox 0.8$ indicate highland regions, while for Al/Si, non-mare regions are $\gtrapprox 1.25$. 
We can derive the following relation to map line ratios to abundances:

\begin{align}
    \frac{\mathrm{CLASS~Mg~wt\%}}{\mathrm{CLASS~Al~wt\%}} &= 0.88 \times \mathcal{R}^\text{Mg}_\text{Al} - 0.34 \label{eq:mgal_class} \\
    \frac{\mathrm{Sample~Mg~wt\%}}{\mathrm{Sample~Al~wt\%}} &= 0.91 \times \mathcal{R}^\text{Mg}_\text{Al} - 0.34  \label{eq:mgal_sample} \\
    \frac{\mathrm{CLASS~Al~wt\%}}{\mathrm{CLASS~Si~wt\%}} &= 0.50 \times  \mathcal{R}^\text{Al}_\text{Si} + 0.09 \label{eq:alsi_class} \\
    \frac{\mathrm{Sample~Al~wt\%}}{\mathrm{Sample~Si~wt\%}} &= 0.53 \times  \mathcal{R}^\text{Al}_\text{Si} - 0.19 \label{eq:alsi_sample}
\end{align}

The conversion enables better utilization of the global high-resolution maps derived here.
\section{Lunar chemistry with elemental ratios}\label{sec:lunar-chemistry}

\begin{figure}[h!]
    \centering
    \includegraphics[width=\linewidth]{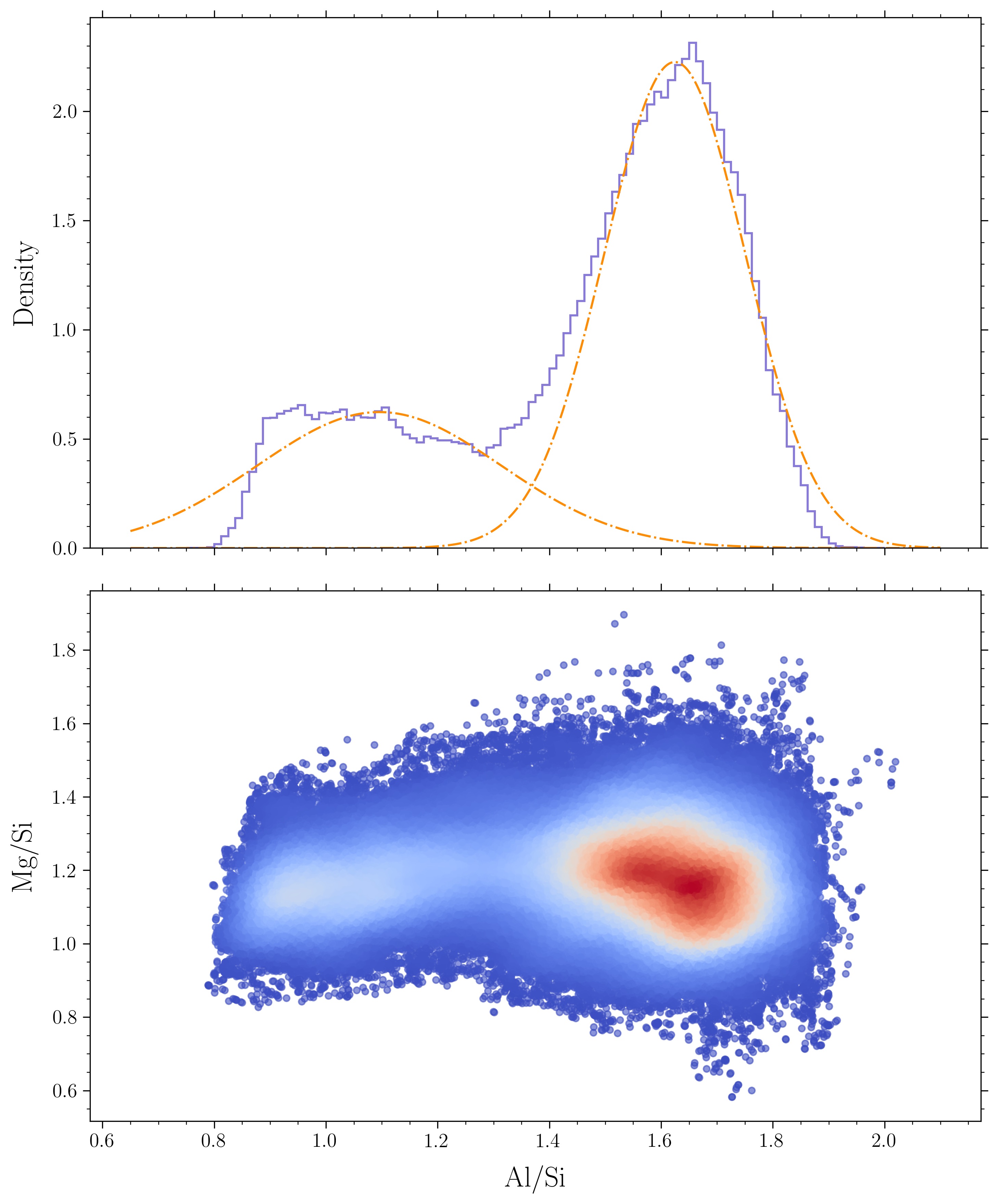}
    \caption{\textit{Top panel:} A histogram of Al/Si line ratios measured over the entire surface shows a clear bimodal distribution, corresponding to the Mare and highland terranes. \textit{Bottom panel:} A two-dimensional distribution of the Mg/Si and Al/Si line ratios also shows a multi-modal distribution.}
    \label{fig:histograms}
\end{figure}

The most powerful combination of ratios, according to our analysis, is the 2D histogram of $\mathcal{R}^\text{Mg}_\text{Si}$ and $\mathcal{R}^\text{Al}_\text{Si}$ (Figure~\ref{fig:histograms}) which easily bifurcates into a region of high Al and low Al, as expected.

\begin{figure*}[htbp]
    \gridline{\fig{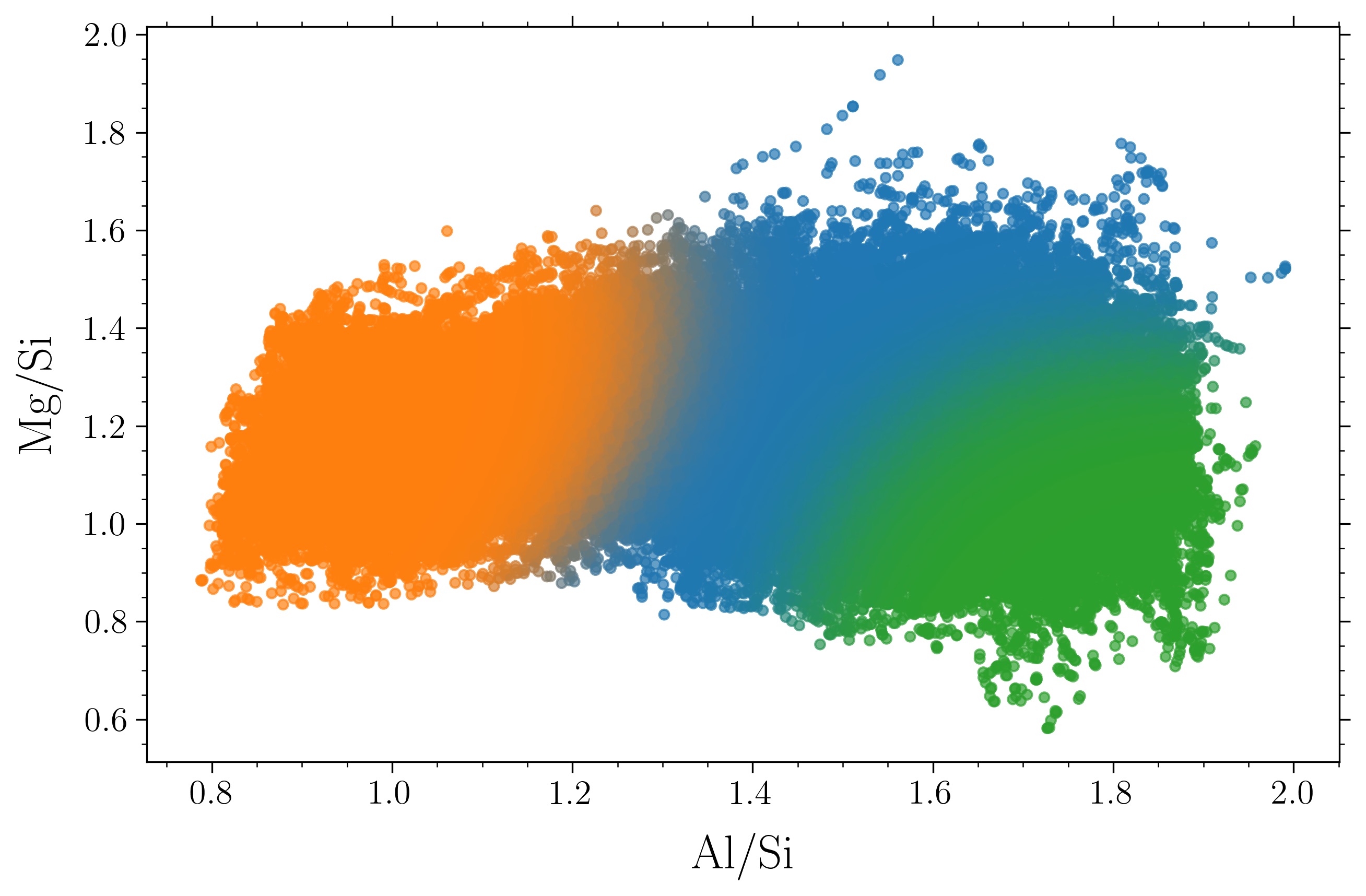}{0.5\textwidth}{}
    \raisebox{17pt}{\fig{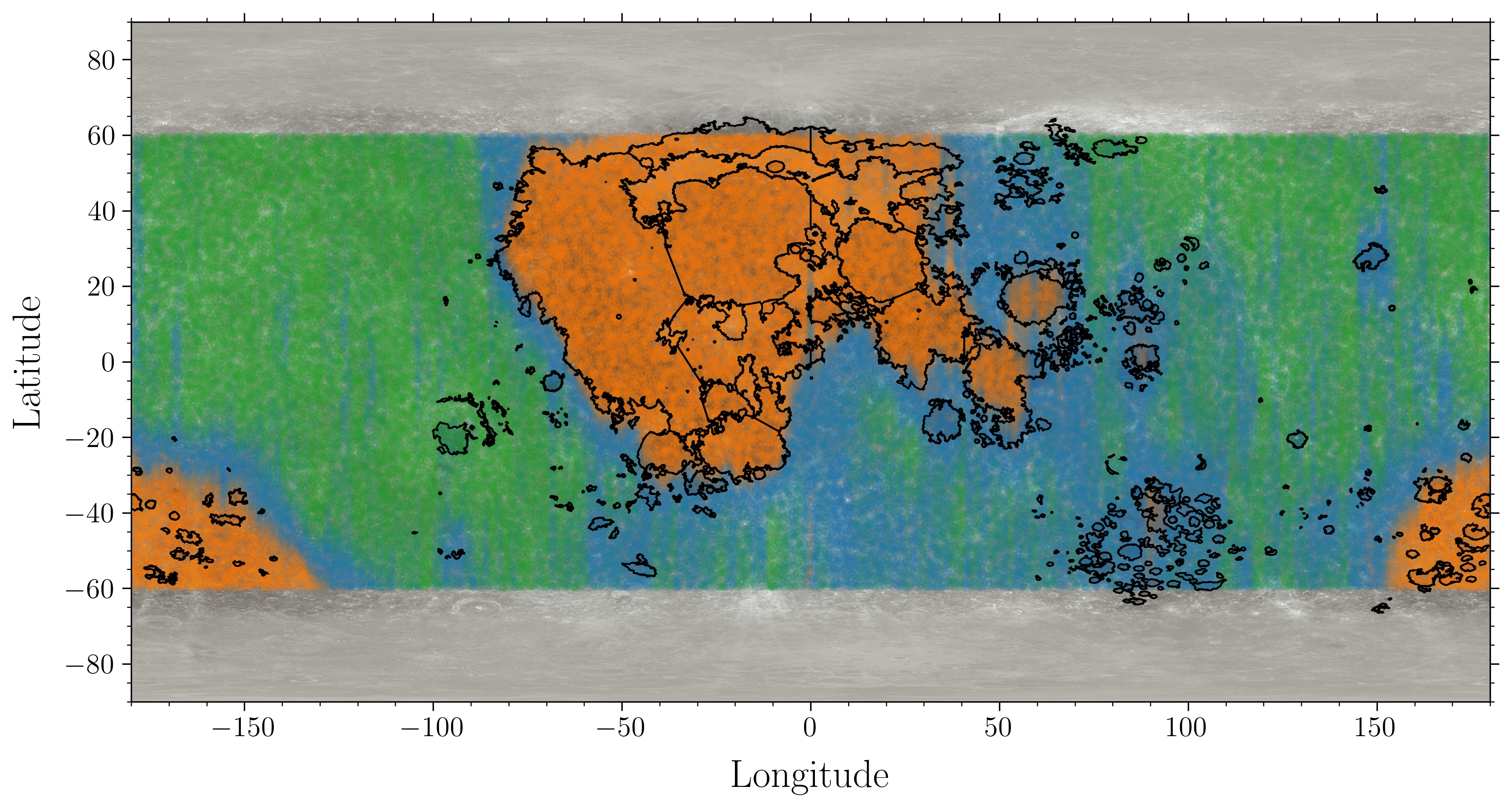}{0.5\textwidth}{}}}
    \vspace{-25pt}
    \gridline{\fig{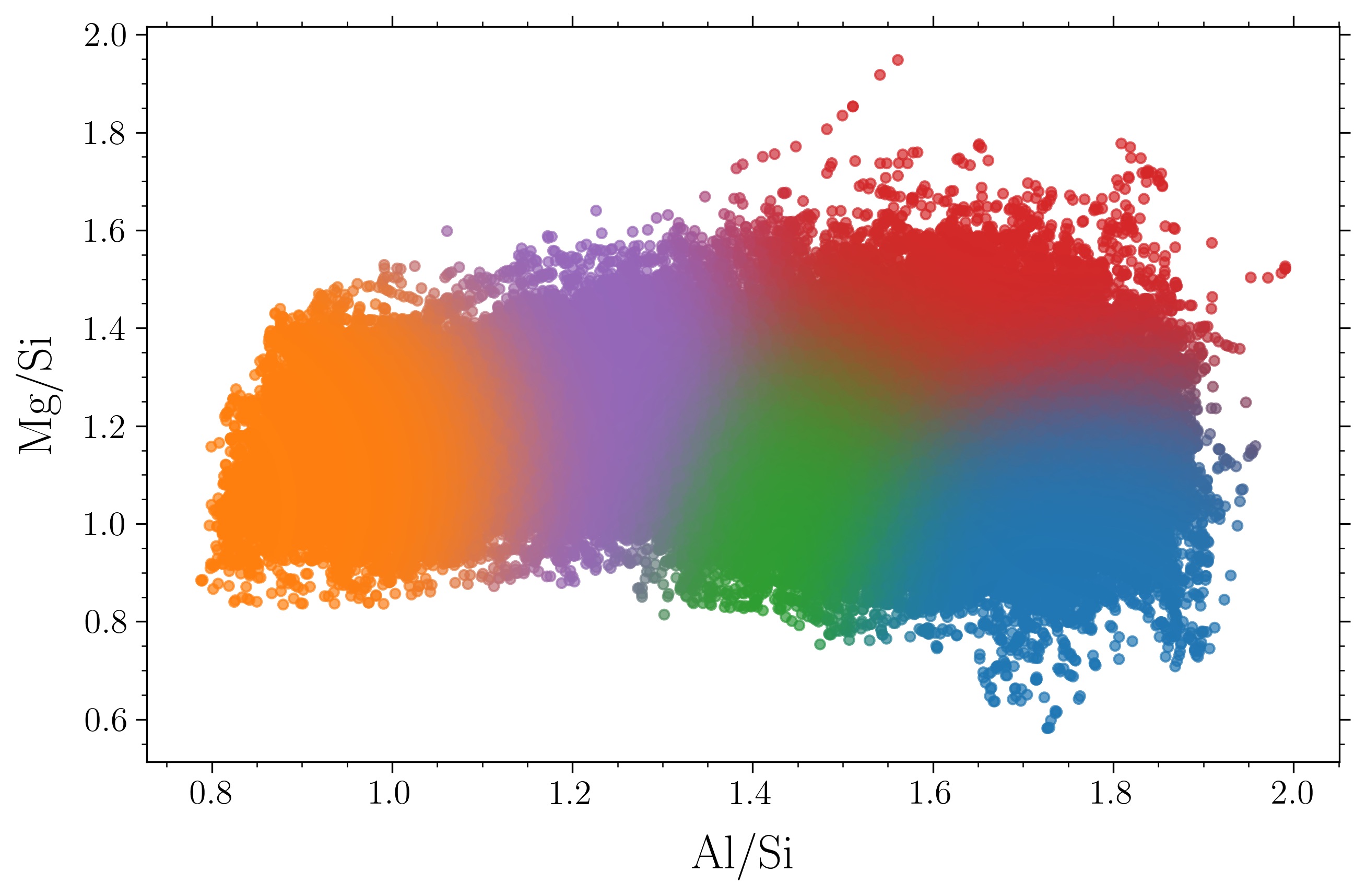}{0.5\textwidth}{}
    \raisebox{17pt}{\fig{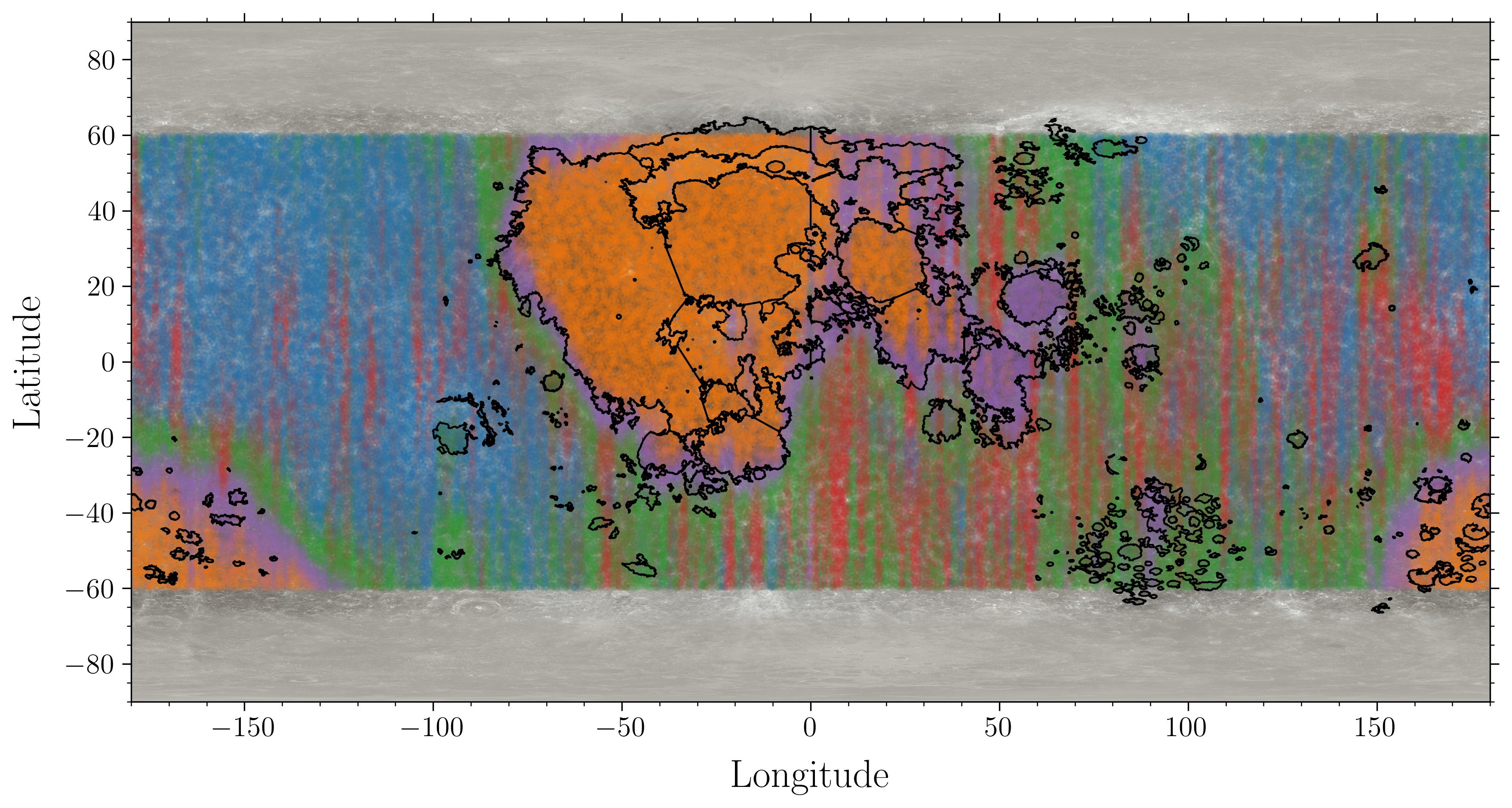}{0.5\textwidth}{}}}
    \vspace{-25pt}
    \gridline{\fig{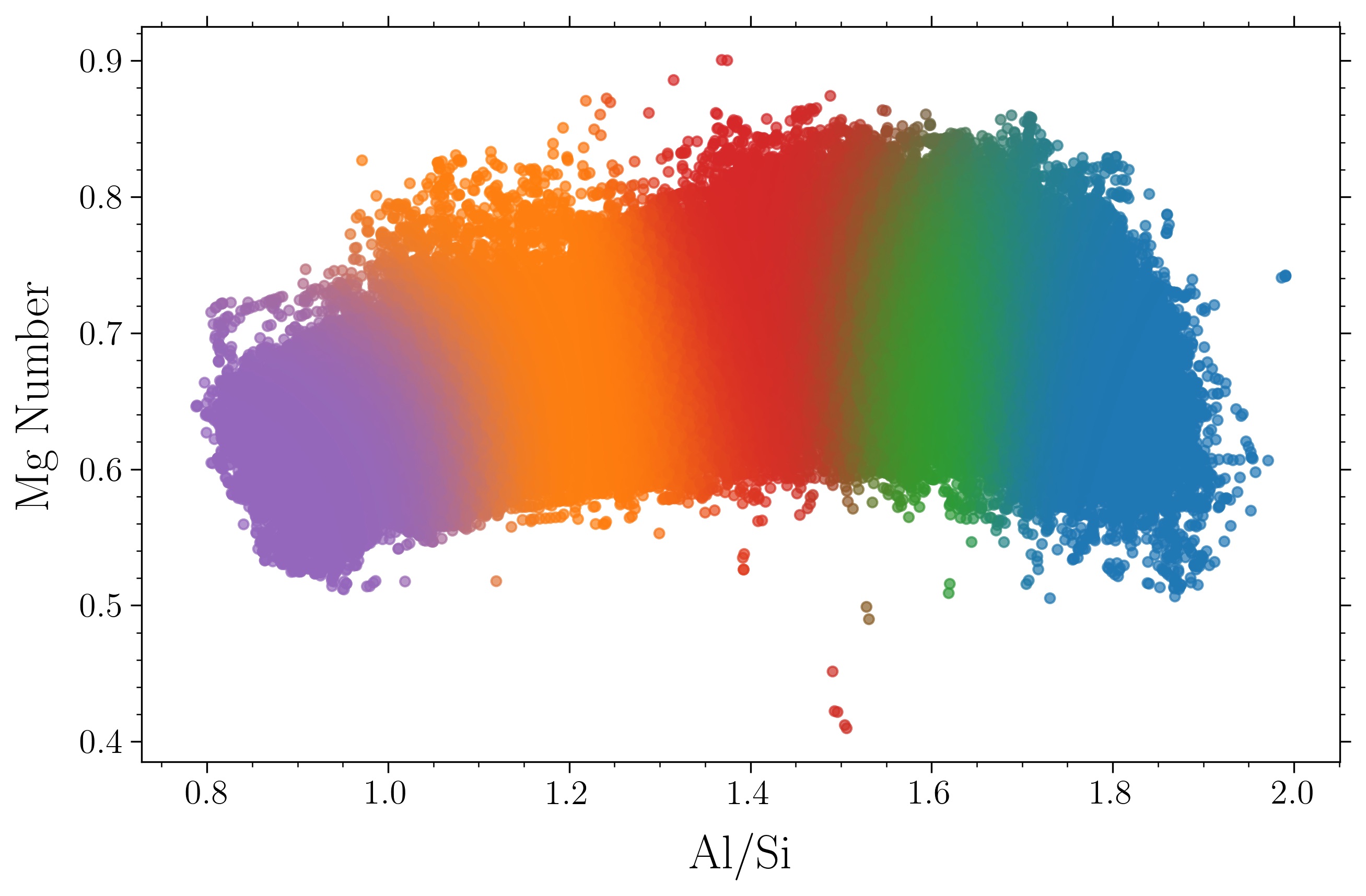}{0.5\textwidth}{}
    \raisebox{17pt}{\fig{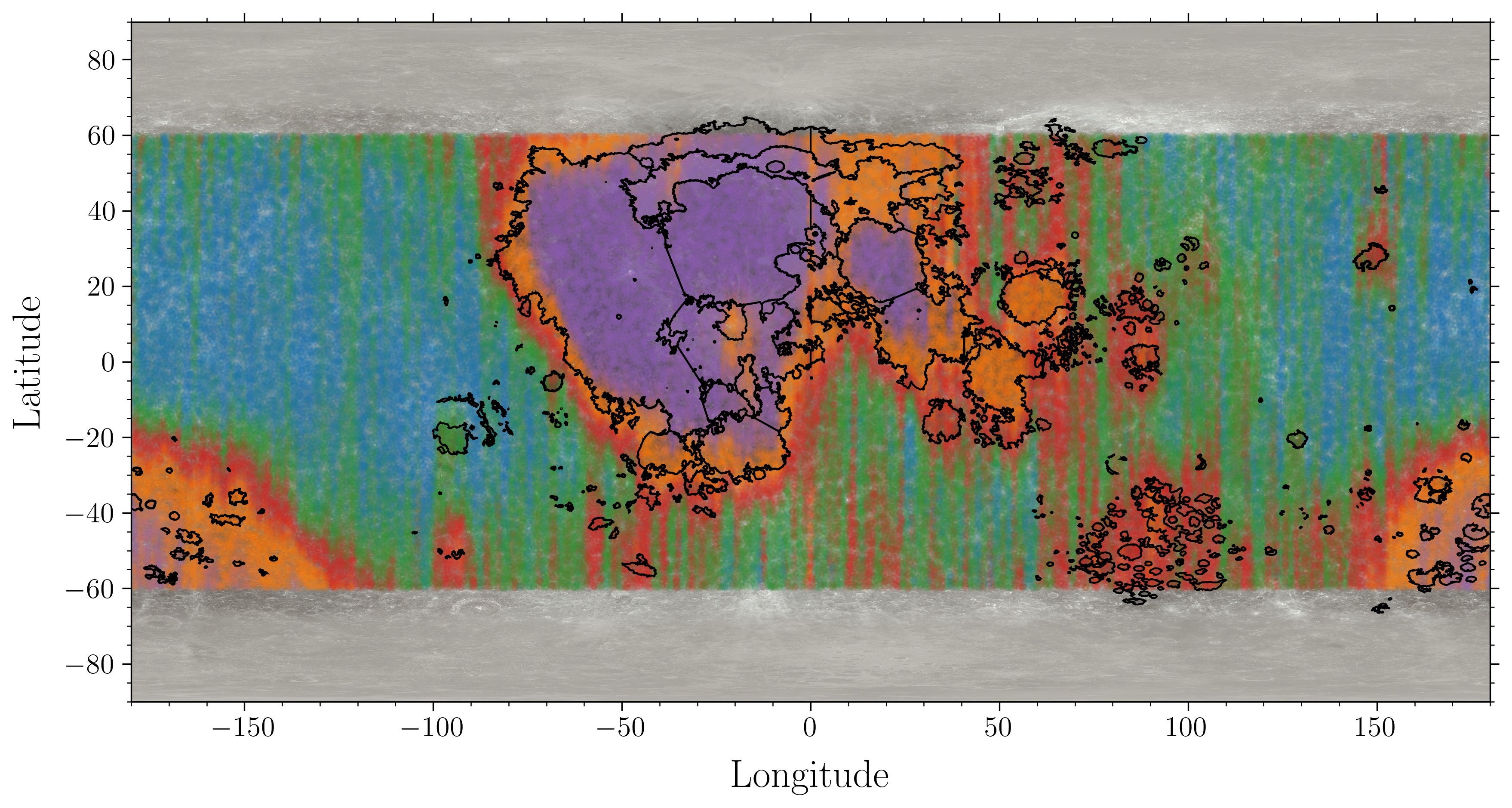}{0.5\textwidth}{}}}
    \caption{Lunar XRF line intensity ratio maps from CLASS data. Black lines are Mare boundaries from LROC data. Mg/Al and Al/Si best represents the distinction between the major lunar terranes. Mg/Si is higher in the mare and SPA. Mg\# proxy value from Mg/Si and Fe/Si line ratios are also associated with distinct lunar terranes.}
    \label{fig:maps3}
\end{figure*}

The high Al/Si data points also show two distinct clusters, which we correspond to selenographically different regions (Figure~\ref{fig:maps3}). We fit the 2D histograms of  $\mathcal{R}^\text{Mg}_\text{Si} \text{ vs }   \mathcal{R}^\text{Al}_\text{Si}$  and $\mathcal{R}^\text{Fe-L}_\text{Si} \text{ vs } \mathcal{R}^\text{Al}_\text{Si}$ using a three-component Gaussian mixture model. The peak of each Gaussian is assigned a different color and the points in between are intermediately colored, representing their probability of belonging to any of the underlying populations assumed. Plotting on the lunar mercator projection with this same color scheme, we can see three regions separated out (1) The Basaltic Lunar Maria (2) South Pole Aitken (SPA) Basin (3) the Feldspathic Highlands terrane.  We apply a five-component model which further distinguishes a low Mg/Si cluster in the north west highland Terranne and a low Mg/Si cluster in the north east. Both models show the intermediate Mg/Si material that outlines the mare and the SPA.

We define an analogous quantity to the Magnesium Number (Mg\#) as $\mathcal R^\text{Mg}_\text{Si}/\left({R^\text{Mg}_\text{Si}+R^\text{Fe}_\text{Si}}\right)$ as shown in Figure \ref{fig:maps3} and apply a five-component Gaussian model. Lower Mg\# ratios appear in the mare and higher in highland terrane with a transition between, as observed in previously published Mg\# maps \citep{WANG2021105360,ZHANG2023115505}. The large basins on the near side belong to the lowest Mg\# cluster suggesting younger lava emplacements. In particular, isolated mare, Mare moscovinse and orientale have a higher Mg\# ratio.

\subsection{Global maps}

\begin{figure*}[hbt!]
    \centering
    \includegraphics[width=0.95\linewidth] {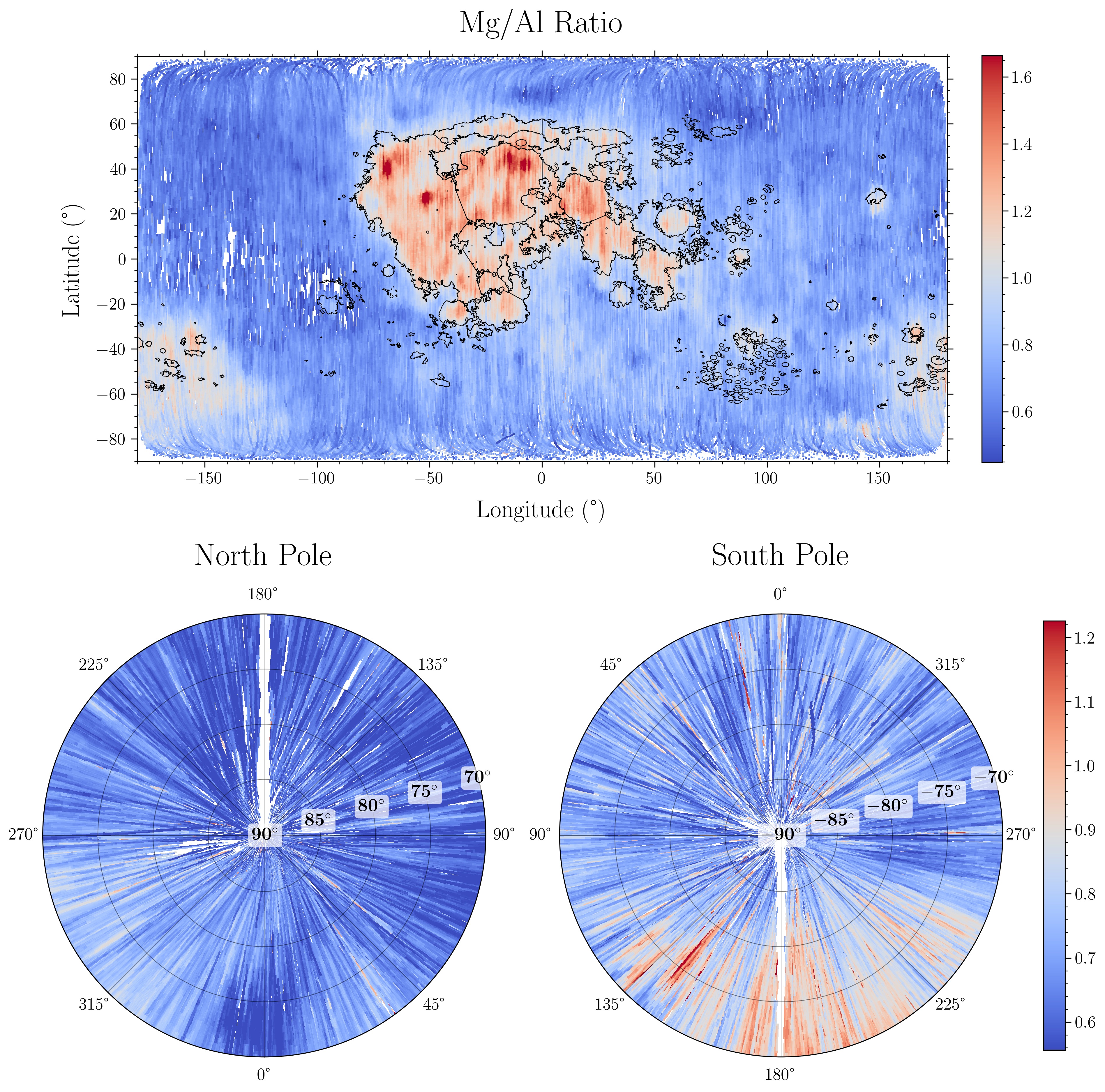}
    \caption{Lunar Mg/Al X-ray intensity maps. \textit{Top panel:} Cylindrical projection. Black lines are Mare boundiares from LROC data. We see a clear distinction in the Mg/Al ratio between Mare and highlands. \textit{Bottom panels:} Maps of the polar regions. The numbers around the periphery are longitudes, while the bold numbers denote latitudes.} 
    \label{fig:Mg_GIS}
\end{figure*}

We have generated global high-resolution maps of major elements with XRF line intensity ratios from CLASS on Chandrayaan-2. The Mg/Al and Al/Si maps are particularly useful to identify geochemically distinct regions that can be used for further detailed studies. The Mg/Al line intensity map is plotted with the contours of the mare basalts in Figure \ref{fig:Mg_GIS}.

The highest Mg/Al regions are associated with the near-side mare basins. Notable regions are the aristarchus plateau and a region close to the boundary of ocenaus procellarum towards the north east. Although the anomalously high Mg/Al ratio ($\approx$ 2 compared to the global average of 0.75) north west of the Aristarchus crater coincides with high olivine abundances on the map derived from the Kaguya multiband imager \citep{2019P&SS..165..230L}, other similar regions do not. These regions are likely to be distinct in mafic content and composition, which requires detailed studies.

The maps in the polar regions (70 - 90$^\circ$) are also shown in the lower panel of Figure \ref{fig:Mg_GIS}.  South Polar region has an average Mg/Al value of $0.73 \pm 0.12$ while it is $0.64 \pm 0.09$ for the North Polar region. In the 85-90$^\circ$ region where most of the future landing zones are located, the Mg/Al, Al/Si and Mg/Si ratios are similar for the polar regions. 

Table \ref{tab:average} gives the average values of Mg/Si, Al/Si and Mg/Al ratios for the polar regions, highlands and mare. Our values are higher than those derived from Chang'e 2 \citep{2016RAA....16....4D} and Apollo XRS \citep{Gloudermans2021}. 

  \begin{table}[h]
    \centering
    \begin{tabular}{|c|c|c|c|c|}
        \hline
        Region & Mg/Si& Al/Si& Mg/Al\\ \hline
        North Polar   & $0.99 \pm 0.11$ & $1.48 \pm 0.31$   & $0.64 \pm 0.09$ \\ \hline
        South Polar   & $1.05 \pm 0.12$   & $1.36 \pm 0.36$   & $0.73 \pm 0.12$ \\ \hline
        Highlands & $1.14 \pm 0.14$ & $1.59 \pm 0.20$ & $0.69 \pm 0.11$ \\ \hline
        Mare & $1.12 \pm 0.14$ & $1.18 \pm 0.17$ & $0.96 \pm 0.22$ \\ \hline
    \end{tabular}
    \caption{Average values of elemental line ratios in the lunar polar regions ($70^\circ-90^\circ$), Highlands and Mare.}
    \label{tab:average}
\end{table}

\section{Summary}
We have developed a processing methodology to detect XRF signals from the CLASS calibrated spectral files and generate high-resolution maps of line intensity ratios. Using five years of CLASS data, global line intensity ratio maps of major elements are generated. The Mg/Al and Al/Si line intensity ratios correlate well with abundance ratios in wt\% from return samples, as well as published CLASS elemental maps.  The Mg/Al maps best represent the geochemical diversity that shows distinct regions that would be of interest for further studies.

\facilities{Chandrayaan-2}

\software{  
            \texttt{Python v3.10.15} \citep{python09},
            \texttt{NumPy} \citep{NumPy20}, 
            \texttt{SciPy} \citep{SciPy20}, 
            \texttt{Astropy} \citep{astropy:2013, astropy:2018, astropy:2022},
            \texttt{Pandas} \citep{mckinney10}, 
            \texttt{Matplotlib} \citep{Hunter07}
        }

\bibliography{sample631}{}
\bibliographystyle{aasjournal}

\end{document}